\documentstyle[prl,aps,twoside,floats,epsf]{revtex}
\begin{document}
\draft
%%%\twocolumn[\hsize\textwidth\columnwidth\hsize\csname
%%%@twocolumnfalse\endcsname
\draft % RevTeX 3.0
\title{
Impurities and Inelastic Processes in Magnetic
Tunnel  Junctions
}
\author{ A.M. Bratkovsky$^{*}$ and J.H. Nickel }

\address{
Hewlett-Packard Laboratories, 3500~Deer~Creek 
Road, Palo Alto, California 94304-1392 }
\date{April 10, 1998}
\maketitle
\begin{abstract}
We have studied tunnel magnetoresistance (TMR) in junctions
with $3d$ ferromagnetic electrodes. 
Previously we predicted that defects in the barrier would
result in reduced effective polarization $P$ of the impurity
assisted current. This is confirmed experimentally in the
present work:
introductions of defects into the barrier 
drastically decreases the TMR.
The degradation of magnetoresistance with bias has also been studied
and shows universal features, attributed to effects of tunneling
assisted by magnons and phonons, whose different role is described.
Details of the bias dependence of the TMR depend on preparation
procedures and well described by the model which includes
assisted tunneling. Non-linear features, seen at low biases,
are related to excitation of bulk modes by tunneling electrons.
The analysis of factors resulting in fall-off of the TMR with bias is 
presented.

\end{abstract}
\pacs{73.40.Gk, 73.40.Rw, 75.70.Pa, 85.70.Kh}
%%%\vskip 2pc ] % end \twocolumn[...]
%\narrowtext

%%%%%%%%%%%%%%%%%%%%%%%%%%%%%%%%%%%%%%%%%%%%%%%%%%%%%%%%%%%%%%%
%%Introduction
%%%%%%%%%%%%%%%%%%%%%%%%%%%%%%%%%%%%%%%%%%%%%%%%%%%%%%%%%%%%%%%
Current through a tunnel structure with ferromagnetic electrodes is
controlled by the mutual orientation of their magnetic moments.
Although this concept was established in early
experiments,\cite{jul,maekawa,tedrow} only recently was it
demonstrated that the
effect can be large at room temperatures.\cite{moodera}
The effect of tunnel magnetoresistance (TMR) is of
fundamental interest and potentially applicable to
magnetic sensors and  memory devices.\cite{tedrow}

According to the standard picture of tunneling, the total resistance
of such junctions depends exponentially on barrier parameters (thickness and
 barrier height). The relative {\em change} in resistance, however, depends
only on the effective polarization of electrons in the
electrodes.\cite{jul,stearns,amb_tunn1}
Two pertinent problems relating to the tunnel magnetoresistance are 
(i) the role of defect/impurity states in the barrier and resonant
tunneling via them, and (ii) the degradation of the TMR with bias.

We shall show that TMR is drastically reduced, as predicted,\cite{amb_tunn1}
by an introduction of impurities into the barrier.
The impurities decrease
polarization of tunneling electrons.
Furthermore, we shall show that the degradation 
of TMR with increasing bias voltage is related to inelastic tunneling
processes involving magnons and phonons, which play opposite roles in TMR.

Magnetoresistance (MR) is defined as the relative change in the junction
conductance with respect to the change of the mutual orientation of spins
from parallel ($G^{\rm P}$)  to antiparallel ($G^{\rm AP}$).
If the effective polarization of bulk tunneling carriers is $P_{\rm FB}$,
then the MR is given by Julliere's formula\cite{jul,amb_tunn1}
\begin{equation}
 {\rm MR} = {G^{\rm P}-G^{\rm AP} \over{ G^{\rm AP} }}=
{2P_{\rm FB}P'_{\rm FB}  \over { 1-P_{\rm FB}P'_{\rm FB}}}.
\label{eq:MR}
\end{equation}
For bulk ferromagnetic electrodes, the values of polarization have
been estimated from low-temperature tunneling experiments.\cite{tedrow}
These values, as well as a straightforward calculation of the TMR due to
direct tunneling between bulk ferromagnetic electrodes,\cite{amb_tunn1}
give a TMR of about~30\%.

The formula (\ref{eq:MR}) 
is generic and applies to physically different processes, 
although the corresponding effective polarizations of these processes
may be very different. 
%%%%%%%%%%%%%%%%%%%%%%%%%%%%%%%%%%%%%%%%%%%%%%%%%%%%%%%%%%%%%%%
%Impurity-assisted tunneling
%%%%%%%%%%%%%%%%%%%%%%%%%%%%%%%%%%%%%%%%%%%%%%%%%%%%%%%%%%%%%%%
The presence of impurity/defect states in the barrier results in
resonant tunneling of electrons.
It dominates over 
direct tunneling  when the impurity density of states
exceeds approximately $10^{17}$cm$^{-3}$eV$^{-1}$.\cite{amb_tunn1} 
The corresponding effective polarization of impurity channels is
significantly lower than the bulk one, and
 it gives a predicted value of about 4\% for TMR.\cite{amb_tunn1}
The much lower polarization is mainly  due to mixing of the tunneling 
electron's wave function with that of non-polarized defect states.
In the case of magnetic impurities (spin-flip centers) 
the TMR will be even smaller. 
As we report below, these
predictions are in good correspondence with our observations.

We have performed a series of experiments to check
the relevance of the afore-mentioned
different physical mechanisms of tunneling. Junctions used were
fabricated with NiFe and/or CoFe electrodes with about 2nm of
plasma oxidized Al to form a barrier.
The resistance change of the NiFe junctions is consistently about 30-32\% with
respect to the lowest resistance of the hysteresis loop (Fig.~1).
Measurements of tunnel resistance as a function of external
field have been performed, including the resistance change in the
field as a function of bias, all at room temperature. An important
result is that the value of TMR does not depend, within the accuracy of our
measurements, on the thickness of the barrier, which has been varied 
between 2.0 to 3.3 nm of alumina. This means that the polarization of
tunneling current is defined mainly by the bulk ferromagnetic
electrodes and not much dependent on the thickness of the clean barrier.

To study the effect of imperfections in the barrier,
the latter has been intentionally modified
to incorporate defects. A uniform distribution 
of impurities across the barrier has been achieved by co-sputtering
Au with Al instead of the usual Al only deposition 
described above. 
Two samples were prepared with 5\% and 10\% Au
incorporated in the Al layer before plasma oxidation.
This leads
to a reduction of TMR down to 17\% and 10\%,
respectively (Fig.~1). We have also prepared
barriers with midlayers of NiFe and Au of thickness $t$. 
Samples with $t_{\rm
midlayer} =0.5$ and 1~nm show zero TMR. The drastic difference between
co-sputtered samples and samples with the metallic midlayer is easy to
understand: only impurities very close to
the center of the barrier are effective scatterers.\cite{amb_tunn1}
Even without spin-flip effects they suppress TMR to a small value, and
the addition of spin mixing due to the exchange interaction 
of tunneling electrons with impurities will quickly reduce it further.
Similar results for midlayers of different $3d$ metals have been 
reported.\cite{mood_def}

%%%%%%%%%%%%%%%%%%%%%%%%%%%%%%%%%%%%%%%%%%%%%%%%%%%%%%%%%%%%%%%%%%%%%%%
%`zero-bias' anomaly
%%%%%%%%%%%%%%%%%%%%%%%%%%%%%%%%%%%%%%%%%%%%%%%%%%%%%%%%%%%%%%%%%%%%%%%
Another important issue is the behavior of TMR under bias conditions.
Generally, if only direct tunneling between ferromagnetic electrodes
were to take place, the TMR would smoothly degrade with bias
following a parabolic law 
[${\rm MR}(V) \approx {\rm MR}(0)-const\times V^2$].\cite{amb_tunn1}
The experimentally observed
degradation is considerably faster at low biases within about 0.03-0.1~V,
with the conductance showing the so-called `zero-bias' 
anomaly.\cite{zerobias}
Our studies of TMR in a series of control junctions
show that the bias dependence of normalized TMR has a rather
universal behavior (Fig.~2). A plausible explanation of this
behavior is inelastic tunneling assisted by emission of bosons
(phonons and magnons).\cite{zerobias} Whereas the role of magnons
obviously includes spin mixing that reduces TMR, there might be a
misconception that phonons are not coupled to spin and, therefore,
play no role in TMR. This is not the case, as
the probability of any tunneling event is proportional to initial and
final densities of states (DOS). Since the DOSs are different 
for majority and minority carriers, {\em any} boson-assisted event would
change TMR. But since phonons do not induce spin mixing, they
{\em increase} the TMR, as opposed to magnons that always
tend to decrease the TMR. In addition, we note that Kondo-like scattering on
magnetic impurities \cite{kondo} (formally third order in exchange
constant) operates at very low biases, if at all,
and usually does not fit experimental $I-V$ curves.\cite{beasley}

Magnon-assisted current in, for example,  parallel configuration 
at bias $V$ and temperature $T$ 
is given by the following expression\cite{amb_tunn2}:
\begin{eqnarray}
I^x_{\rm P}(V,T) &=& 
{2\pi e\over \hbar}\sum_{{\bf q}\alpha}X^\alpha
\biggl(
g^L_\uparrow g^R_\downarrow \times~(eV+\omega)
\left[
{N_\omega\over {1-\exp(-\beta(eV+\omega))}}
+{{N_\omega+1}\over{ 1-\exp(\beta(eV+\omega))}}
\right]\nonumber\\
&+&g^L_\downarrow g^R_\uparrow \times~(eV-\omega)
\left[
{{N_\omega+1}\over{1-\exp(-\beta(eV-\omega))}}
+{N_\omega\over{1-\exp(\beta(eV-\omega))}}\right]
\biggr),
\label{eq:IxP}
\end{eqnarray}
where $N_\omega = \left[\exp(\beta\omega)-1\right]^{-1}$,
$\beta=1/T$, 
$\omega = \omega^\alpha_{\bf q}$ and 
$X^\alpha$ is the incoherent tunnel exchange (electron-magnon) vertex
with all momenta parallel to the barrier integrated out, 
$g^{L(R)}$ marks
the corresponding electron density of states on the left (right)
electrode,   and $\alpha=L, R$. Analogous expression can be easily
obtained for the AP configuration.
As follows from Eq.~(\ref{eq:IxP}), the  magnon-assisted 
inelastic tunneling current at $T=0$ is
\begin{eqnarray}
I^x_{\rm P} &=& {2\pi e\over \hbar}\sum_\alpha X^\alpha 
g^L_\downarrow g^R_\uparrow
\int d\omega 
\rho^{mag}_{\alpha}(\omega) (eV-\omega)\theta(eV-\omega),\nonumber\\
I^x_{\rm AP} &=& {2\pi e\over \hbar}\biggl[
X^R g^L_\uparrow g^R_\uparrow
\int d\omega \rho^{mag}_{R}(\omega) (eV-\omega)\theta(eV-\omega)\nonumber\\
&+&
X^L g^L_\downarrow g^R_\downarrow
\int d\omega \rho^{mag}_{L}(\omega) (eV-\omega)\theta(eV-\omega)\biggr],
\label{eq:Ix0}
\end{eqnarray}
where 
$\rho_\alpha^{mag}(\omega)$ is the magnon density of states
that has the
general form $\rho_\alpha^{mag}(\omega)
=(\nu+1)\omega^{\nu}/\omega_0^{\nu+1}$, where the exponent
$\nu$ depends on the type of spectrum and may be determined from a fit
to experimental data, 
$\omega_0$ is the maximum magnon frequency, 
and  $\theta(x)$ is the step function.

For phonon-assisted current at $T=0$ we obtain 
\begin{eqnarray}
I^{ph}_{\rm P} &=& {2\pi e\over \hbar}\sum_{a\alpha} g^L_a g^R_a
\int d\omega  \rho^{ph}_\alpha(\omega) \nonumber\\
&&\times P^\alpha(\omega)(eV-\omega)\theta(eV-\omega),\\
I^{ph}_{\rm AP} &=& {2\pi e\over \hbar}\sum_{a\alpha} g^L_a g^R_{-a}
\int d\omega  \rho^{ph}_\alpha(\omega) \nonumber\\
&&\times P^\alpha(\omega)(eV-\omega)\theta(eV-\omega),
\label{eq:Ip0}
\end{eqnarray}
where $P(\omega)$ is the phonon vertex proportional to a deformation
potential,
$P(\omega)/X = \gamma \omega/\omega_D$, $\gamma$ is a constant,
$\omega_D$ is the Debye frequency, $a$ is the spin index, 
$\rho_\alpha^{ph}$ is the phonon DOS,
$\alpha$ here marks electrodes and the barrier.

The elastic and inelastic contributions together will define the total
junction conductance $G=G(V,T)$ as a function of the bias $V$ and
temperature $T$.
We find that the inelastic contributions from magnons and phonons
(\ref{eq:Ix0})-(\ref{eq:Ip0}), respectively, grow as 
$G^x(V,0) \propto (|eV|/\omega_0)^{\nu+1}$ and 
$G^{ph}(V,0) \propto (eV/\omega_D)^4$
at low biases. 
These contributions saturate at higher biases: $G^x(V,0) \propto
1-\frac{\nu+1}{\nu+2}\frac{\omega_0}{|eV|}$ at $|eV|>\omega_0$;
$G^{ph}(V,0) \propto 
1-\frac{4}{5}\frac{\omega_D}{|eV|}$ at $|eV|>\omega_D$.
This behavior would lead to sharp features in the
$I-V$ curves on a scale of 30-100 mV (Fig.~\ref{fig:fit}).

To highlight the opposite effects of phonons and magnons
on the TMR, let us consider the case of the same electrode materials and
denote $D=g_\uparrow$ and $d=g_\downarrow$. We see that
$G^x_{\rm P}(V,0) - G^x_{\rm AP}(V,0) 
\propto - (D-d)^2(|eV|/\omega_0)^{\nu+1}<0$, whereas
$G^{ph}_{\rm P}(V,0) - G^{ph}_{\rm AP}(V,0) 
\propto +(D-d)^2(eV/\omega_D)^4>0 $, i.e.
spin-mixing due to magnons {\em decreases} the TMR, whereas phonons tend to
reduce the negative effect of magnon emission on the 
magnetoresistance.

Finite temperature gives contributions
of the same respective sign as written above.
For magnons: 
$G^x_{\rm P}(0,T) - G^x_{\rm AP}(0,T) 
\propto - (D-d)^2 (-TdM/dT)<0$, 
where $M=M(T)$ is the magnetic moment of the electrode at a given 
temperature $T$. The phonon contribution is given by a standard 
Debye integral with the following results:
$G^{ph}_{\rm P}(0,T) - G^{ph}_{\rm AP}(0,T) 
\propto +(D-d)^2(T/\omega_D)^4>0 $ at $T\ll\omega_D$, and 
$G^{ph}_{\rm P}(0,T) - G^{ph}_{\rm AP}(0,T) \propto
+(D-d)^2(T/\omega_D)$
 at $T\gtrsim \omega_D$. %\cite{amb_tunn2} 
It is worth mentioning that the magnon excitations 
are usually cut off by, for instance, the anisotropy
energy $K_{\rm an}$ at some $\omega_c$. Therefore, at
low temperatures  the conductance 
at small biases will be almost constant.

The experimental data in Fig.~2(a)
has been fitted by expressions (3)-(5) with the inclusion of the bias
dependence of direct tunneling current. The figure shows 
the bias dependence of the direct tunnel current (broken line)
is almost parabolic and does not replicate the data. The inclusion of
inelastic events, however, produces a very good fit
to the experimental data. Moreover, we extract an exponent 
for the magnon DOS which happens to be close to $\nu \approx
\frac{1}{2}$.
This square-root law is characteristic of usual bulk magnons
which, therefore, are of major importance in the `zero bias' anomaly of
the TMR.
Two-dimensional magnons, if they were important, would result
in a linear cusp at zero bias ($\nu=0$), which has not 
generally been observed.\cite{zhang}
The present fit shows that the bulk, rather than surface magnons, are
responsible for the shape of the $I-V$ curves at  low biases (Fig.~2).

The effect of preparation scheme and impurities is further illustrated
by Fig.~2(b). There we compare the data for plasma oxidized, 
radio frequency (rf) sputtered Co/AlO$_x$/NiFe samples, 
together with NiFe/AlO$_x$/NiFe
with 5\% and 10\% of co-sputtered Au in the barrier. In order to
understand this data for the bias dependence of TMR, it is helpful 
to cast (3)-(5) in the form
$ {\rm MR}(V)=[{\rm MR}_0(V)-p_x B_1(V)]/[1+p_x B_2(V)]$, 
where ${\rm MR}_0(V)$ is the tunnel magnetoresistance due to direct
tunneling, $B_1$ and $B_2$ are smooth positive functions of the voltage bias,
and $ p_x  \propto  X/G^0_{\rm AP} $ is the parameter inversely
proportional to an elastic conductance for the antiparallel configuration.
Consequently, the parameter $p_x$ is larger for more resistive
junctions. 
Our  rf sputtered junction were five times more resistive
than the plasma oxidized ones, resulting in a steeper decrease of 
their TMR with bias [Fig.~2(b)]. We find that in all cases the data
is clearly non-linear at low biases, supporting our conclusion
about the dominating role of the {\em bulk} magnon excitations 
in the voltage dependence of the tunnel 
magnetoresistance (Fig.~2).

%%%%%%%%%%%%%%%%%%%%%%%
%%conclusions
%%%%%%%%%%%%%%%%%%%%%%%
The present results demonstrate over 30\% 
completely stable resistance change in magnetic tunnel junctions at
room temperature.
Presence of impurities, defect states in the barrier, significantly reduces 
the magnetoresistance. We expect that any inhomogeneity in the barrier
would lead to similar results.
The rf sputtered junctions are likely
to have a larger density of  defect states in the barrier,
and, consequently, smaller tunnel magnetoresistance
as compared to plasma oxidized junctions. 
It is demonstrated, that
the inelastic processes are  responsible for the unusual shape of
the $I-V$ curves at low biases, and their temperature behavior,
which is also affected by impurity-assisted tunneling. 
The fall-off of tunnel magnetoresistance is faster in  more resistive
rf sputtered junctions.
Trends and details of the bias dependence of the TMR are well
described by the present theoretical model.
Analysis of the
data suggests that the anomalies are due to emission of bulk magnons
and phonons. Further study is required to demonstrate to what
extent these characteristics can be controlled to improve
performance of the junctions.

\newpage
%%%%%%%%%%%%%%%%%%%%%%%%%%%%%%%%%%%%%%%%%%%%%%%%%%%%%%%%%%%%%%%%%%%%
\begin{figure}[t]
\epsfxsize=3.4in
\epsffile{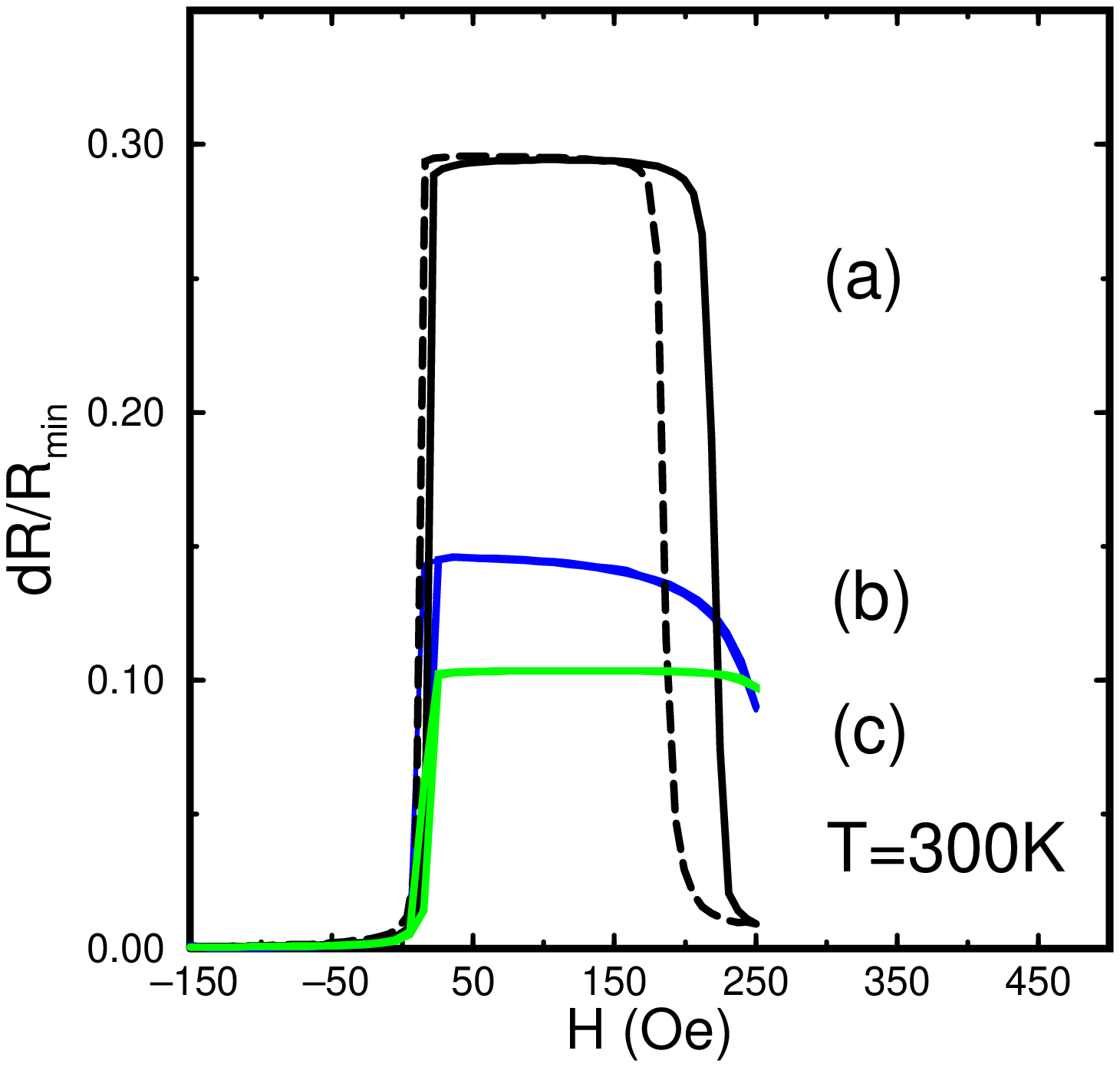}
\caption{\footnotesize
Room temperature magnetoresistance 
($dR/R_{min}$) curves for tunnel junctions
prepared as described in text (thicknesses in nm): 
%%%
(a) junction NiFe/2AlO$_x$/NiFe,
area 1800~$\mu m^2$, $R_{min} = 10~k\Omega$;
%%%
(b) junction with co-sputtered Au in a barrier
NiFe/1.9AlO$_x$0.1Au/NiFe,
area 1800~$\mu m^2$, $R_{min} = 33.5~k\Omega$.
%%%
(c) junction with co-sputtered Au in a barrier
NiFe/1.8AlO$_x$0.2Au/NiFe,
area 1800~$\mu m^2$, $R_{min} = 17~k\Omega$.
Solid curves: forward sweep in field, broken curve: backward sweep. 
Switching field $H_{sw}=250$~Oe is due to exchange biasing by MnFe. 
Samples with NiFe midlayer (not shown) demonstrate zero 
magnetoresistance. 
%%%
\label{fig:tmr3}
}
\end{figure}
%%%%%%%%%%%%%%%%%%%%%%%%%%%%%%%%%%%%%%%%%%%%%%%%%%%%%%%%%%%%%%%%%%%%

\newpage
%%%%%%%%%%%%%%%%%%%%%%%%%%%%%%%%%%%%%%%%%%%%%%%%%%%%%%%%%%%%%%%%%%%%
%	FIGURE FIT
%___________________________________________________________________
\begin{figure}[t]
\epsfxsize=3.4in
\epsffile{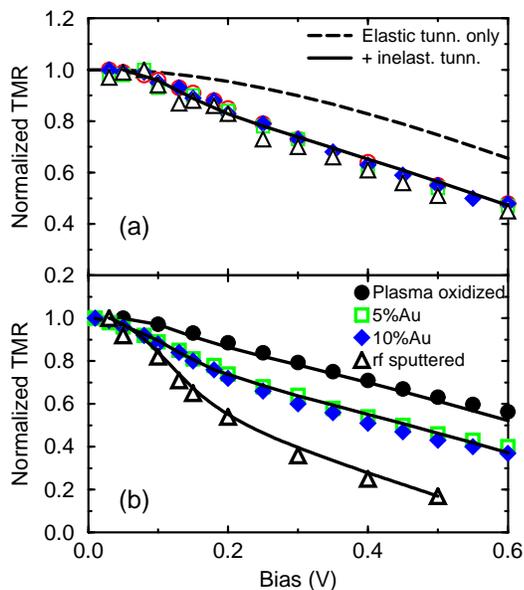}
\caption{\footnotesize
%%%
(a) Experimental data for the magnetoresistance of CoFe/Al$_2$O$_3$/NiFe 
tunnel junctions. The junctions were prepared similar to those
described in text, 
with the bottom electrode containing only Co or CoFe, 
with $dR/R_{min}$=0.10.
The fit includes elastic and inelastic (assisted 
by magnons and phonons) tunneling. 
The fit gives a magnon DOS very close to $\propto \omega^{0.5}$, which is 
characteristic of the bulk spectrum.
%%%
(b) Experimental data for the magnetoresistance of plasma oxidized
and rf sputtered  CoFe/Al$_2$O$_3$/NiFe tunnel junctions,
and NiFe/Al$_2$O$_3$/NiFe junctions with 5\% and 10\% of co-sputtered
Au. In all cases the bias dependence at low voltages in non-linear,
indicating the emission of the bulk magnons. Exponent in magnon
spectrum was again close to the bulk one $\omega^{0.5}$, 
parameter $p_x$ varied following changes in junction conductance 
(see text).
\label{fig:fit}
}
\end{figure}
%%%%%%%%%%%%%%%%%%%%%%%%%%%%%%%%%%%%%%%%%%%%%%%%%%%%%%%%%%%%%%%%%%%%

\end{document}